\documentclass[12pt]{article}
\usepackage[english]{babel}
\usepackage[dvips]{graphicx}
\usepackage{exscale}
\usepackage{epsfig}
\usepackage{amsmath}
\evensidemargin=-7mm \addtocounter{page}{0} \tolerance=5000

\newcommand{\si}{\mbox{{\boldmath$\sigma$}}}
\newcommand{\Si}{\mbox{{\boldmath$\Sigma$}}}
\newcommand{\li}{\mbox{${\bf l}$}}
\newcommand{\vp}{\mbox{${\bf p}$}}
\newcommand{\vk}{\mbox{${\bf k}$}}
\newcommand{\ve}{\mbox{${\bf e}$}}
\newcommand{\gr}{\mbox{{\boldmath$\nabla$}}}
\newcommand{\vr}{\mbox{${\bf r}$}}
\newcommand{\vn}{\mbox{${\bf n}$}}
\newcommand{\vI}{\mbox{${\bf I}$}}
\begin{document}
\begin{titlepage}
\begin{center}
{\Large \bf  Parity nonconservation effects in the photodesintegration of polarized
deuterons}
\end{center}

\begin{center}
 R.V. Korkin\\ Budker
Institute of Nuclear Physics\\ and Novosibirsk University
 630090 Novosibirsk, Russia.\\
\end{center}

\begin{abstract}
P-odd correlations in the deuteron photodesintegration are considered. The $\pi$-meson
 exchange is not operative in the case of unpolarized
deuterons. For polarized deuterons a P-odd correlation due to the $\pi$-meson exchange
is about $3 \times 10^{-9}$. Short-distance P-odd contributions exceed essentially
than the contribution of the $\pi$-meson exchange.
\end{abstract}



\vspace{7cm}

\end{titlepage}

\section{Introduction}

  In the past few years discussions about measuring of cross-section asymmetry in the
  deuteron photodisintegration  with polarized photons becomes very popular according to the
  experimental progress: creating intense  sources of polarized photons, electrons, and
neutrons.
   On the other side, theoretical treatments of deuteron are relatively
   reliable due to the small bound energy.
 One may hope that they will resolve a contradiction between
experiments with $^{133}$Cs and other nuclear experiments which exists at
present~[1..6].

 The theoretical studies parity nonconservation effects in the deuteron were started
 in [7-11]. The electron-deuteron scattering was investigated in [11-13]. However in this process  the
 effect of the nuclear parity violation is masked by neutral currents. Numerical
 estimates of this effect were made in [14-16].
Desintegration of deuterons by polarized photons were considered in [16-21].
 The point is that
 the nuclear anapole moment (AM) of $^{133}$Cs was recently discovered and measured
with good accuracy in atomic experiment~\cite{wi}. The result of this experiment is in
a reasonable quantitative agreement with the theoretical predictions, starting
with~\cite{fk,fks}, if the so-called ``best values''~\cite{ddh} are chosen for the
parameters of P-odd nuclear forces.
  But as it is shown in the paper \cite{korkh}, the contribution to the total cross-section
  asymmetry due to the $\pi$-meson exchange vanishes. The contribution
  of vector meson exchange is essential, and gives the maximal magnitude for asymmetry
    $10^{-7}$. However the vector meson ($\rho,\omega$) exchange has the typical range
    $r_{V}\sim 1/m_{V}\sim 0.25 fm$, that is less than the nucleon size $0.86 fm$. So because of
    unreliable theoretical estimates this contribution is not so interesting.
   Thus it is interesting to consider polarized deuterons photodisintegration.
  We make all of calculations in the potential model.
   It is legitimate since the typical range of interaction $1/m_{\pi}$ is essential
    more than the experimental nucleon size.

 We use in present paper the deuteron wave function in the zero range approximation.

\begin{equation}
\psi_{d}^{\sigma}=\sqrt{\frac{\kappa}{2\pi\,(1-\kappa\,r_{t})}}\frac{e^{-\kappa\,r}}{r}
\chi^{\sigma},
\end{equation}
where $\chi^{\sigma}$ is a spin wave function.

It is, strictly speaking, inconsistent, but this function fails only at the small
range and  give us the correct behavior if the radius is much more than the effective
radius of the triplet state $r>>r_{t}$. Matrix element of the  electric dipole
transition does not depend on function behavior at small range and we may use the zero
range approximation for its calculation. As to the magnetic dipole transition or for
the weak potential matrix element the situation is worse because of a sensitivity to
the small range contribution their operators. Let us mention that this function does
not satisfy to the normalization condition with the correction factor
$\sqrt{1-\kappa\,r_{t}}$ and if we wish to make calculations sensitive to the small
range with this function we must modify it for the right normalization. It is easy to
understand that we must neglect $\sqrt{1-\kappa\,r_{t}}$ \cite{blp}.

\section{Amplitudes calculation}

   The operators of electric dipole and magnetic dipole transitions are \cite{blp}
\[
  H_{E1}=-i\,e\,\sqrt{2\,\pi\,\omega}\,\mbox{${\bf n}$}\ve\,\frac{r}{2},
\]
\begin{equation}
  H_{M1}=-i\,\frac{e}{2\,m}\,\sqrt{2\,\pi\,\omega}\,(\mu_{p}\si_{p}+\mu_{n}\si_{n}+
\frac{\li}{2})\,\frac{\mbox{${\bf k}$}\times\ve}{k}.
\end{equation}

 Where $r_{p}$ and $l_{p}$ are proton coordinate and momentum, $\ve$ and $\vk$ are
 polarization and momentum of photon.
   Now we may calculate the amplitude of the regular electric dipole transition in the deuteron
    photodesintegration. The initial state is the deuteron wave function $\psi_{d}$. It is
    obviously that final state is continuous $p$-wave. Further,
     we may write in amplitude a plane
    wave instead of $p$-wave, since $E1$-transition
    selects by itself only $p$-state from the plane wave.

Then the amplitude is

\begin{equation}
  A_{E1}=-2\,e\,\sqrt{2\,\pi\,\omega}\sqrt{2\,\pi\,\kappa}\,\frac{\vp\,\ve}{\sqrt{1-\kappa\,r_{t}}(\kappa^2+p^2)^2}.
\end{equation}

 Due to the weak parity nonconservation interaction, the initial deuteron $s$-wave has mixture
  of the $p$-wave, and the final $p$-wave has mixture of the $s$-wave. Thus mixing $M1$ transitions
  via intermediate states becomes possible.

 Parity nonconservation $\pi$-meson exchange potential may be written in the following
 form \cite{kk}

\begin{equation}
V=-i\frac{g\overline{g}}{4\,\pi\,m}(\si_{p}+\si_{n})\gr\frac{e^{-m_{\pi}\,r}}{r}.
\end{equation}

 Then we may write the total amplitude of transition

\[
  A=\langle\vp,\sigma'\mid H_{E1}\mid\psi_{d}, \sigma\rangle+
   \sum_{n}{\frac{\langle\vp,\sigma'\mid H_{M1}\mid n
\rangle\langle n\mid V \mid \psi_{d}, \sigma\rangle}{E_d-E_{n}}}
\]
\begin{eqnarray}\label{Atotal}
+\sum_{n}{\frac{\langle\vp,\sigma'\mid V\mid n \rangle\langle n\mid H_{M1}\mid
\psi_{d}, \sigma\rangle}{E_p-E_{n}}}.
\end{eqnarray}

Let us consider the second expression of the (\ref{Atotal}) and calculate
 transition from the deuteron into an intermediate state due to the weak interaction.
As it was mentioned above the $p$ state my be written as a plane wave

\[
\psi(p)=e^{i\,{\bf p}\,{\bf r}}\,\mid\chi^{\sigma}\rangle.
\]

Then this transition is

\begin{equation}
\langle\vp,\sigma''\mid
V\,\mid\psi_{d}\rangle=-i\frac{g\overline{g}}{4\,\pi\,m}\langle\chi^{\sigma''}\mid
(\si_{p}+\si_{n})\mid \chi^{\sigma}\rangle\sqrt{\frac{\kappa}{2\,\pi}}\int e^{-i\,{\bf
k}\,{\bf r}}\gr(\frac{e^{-m_{\pi}\,r}}{r}) \frac{e^{-\kappa\,r}}{r} d\vr.
\end{equation}

  Simple calculations leads to the following
expression for the weak transition

\begin{equation}
\sqrt{\frac{\kappa}{2\,\pi}}\frac{g\,\overline{g}}{k\,m}\langle\chi^{\sigma''}\mid
\vp\,(\si_{p}-\si_{n}) \mid \chi^{\sigma}\rangle f(p).
\end{equation}

 Here $f(p)$ is

 \begin{equation}
f(p)=\frac{1}{2}\left[\frac{m_{\pi}-\kappa}{p}\left(1-\frac{m_{\pi}+\kappa}{p}\arctan\frac{p}{m_{\pi}+\kappa}\right)+\arctan\frac{p}{m_{\pi}+\kappa}\right].
 \end{equation}

 Now one may consider $M1$ transition in the second term (\ref{Atotal})

 \[
 \langle\vp,\chi^{\sigma'}\mid H_{M1}
 \mid \vk,\chi^{\sigma''}\rangle=\,-i\,(2\,\pi)^3\,\frac{e}{2\,m}\sqrt{2\,\pi\,\omega}
 [\vn_{k}\times\ve]
 \]
\begin{equation}
 \langle\sigma'\mid(\mu_{p}\si_{p}+\mu_{n}\si_{n}-i\frac{\vp\times\gr_{p}}{2})\mid\sigma''\rangle
 \delta(\vk-\vp).
\end{equation}

 Then the second term in after summarize over all
of intermediate states (momentum and spin) give us the following expression

\begin{equation}\label{secter}
i\frac{e\,g\,\overline{g}\,f(p)\,\sqrt{2\pi\,\omega}}{2\,p\,m\,(\kappa^2+p^2)}\sqrt{\frac{\kappa}{2\,\pi}}
[\vn_{k}\times\ve]
\langle\chi^{\sigma'}\mid(\mu_{p}\si_{p}+\mu_{n}\si_{n}-\frac{i\,[\vp\times\,\gr_{p}]}
{2})\vp(\si_{p}+\si_{n})\mid\chi^{\sigma}\rangle.
\end{equation}

 Here we used the known completeness relation

 \[
 \sum_{\sigma}\mid\chi^{\sigma}\rangle\langle\chi^{\sigma}\mid\,=\,1
 \]
 to summarize over all of intermediate spin states.

Because of the orthogonality of radial $s$-functions in the last term of the formula
(\ref{Atotal}) intermediate state $n$ must be only the deuteron state. That's why the
angular momentum operator is not operative here. After lengthy calculations we obtain
the result

\begin{equation}
-i\frac{g\,\overline{g}\,f(p)\,\sqrt{2\pi\,\omega}}{2\,p\,m\,(\kappa^2+p^2)}\sqrt{\frac{\kappa}{2\,\pi}}
\left[\vn_{k}\times\ve\right]\langle\chi^{\sigma'}\mid\vp(\si_{p}+\si_{n})
(\mu_{p}\si_{p}+\mu_{n}\si_{n})\mid\chi^{\sigma}\rangle
\end{equation}

Then for the total amplitude after summarize of two terms we find

\[
  A=-2\,e\,\sqrt{2\,\pi\,\omega}\sqrt{2\,\pi\,\kappa}\frac{\vp\,\ve}{\sqrt{(1-\kappa\,r_{t})}(\kappa^2+p^2)^2}\delta_{\sigma\sigma'}
\]

\[
+i\frac{e\,g\,\overline{g}\,f(p)\,
\sqrt{2\pi\,\omega}}{2\,p\,m\,(\kappa^2+p^2)}\sqrt{\frac{\kappa}{2\,\pi}}
\left[\vn_{k}\times\ve\right] \langle\chi^{\sigma'}\mid
(\mu_{p}\si_{p}+\mu_{n}\si_{n})\,\,\vp(\si_{p}+\si_{n})
-\vp(\si_{p}+\si_{n})\,\,(\mu_{p}\si_{p}+\mu_{n}\si_{n})\mid\chi^{\sigma}\rangle
\]

\begin{eqnarray}
+\frac{e\,g\,\overline{g}\,f(p)\,
\sqrt{2\pi\,\omega}}{2\,p\,m\,(\kappa^2+p^2)}\sqrt{\frac{\kappa}{2\,\pi}}
\left[\vn_{k}\times\ve\right] \langle\chi^{\sigma'}\mid
\frac{\left[\vp\times\,\gr_{p}\right]} {2}\vp(\si_{p}+\si_{n})\mid\chi^{\sigma}\rangle
\end{eqnarray}

\section{Differential and total cross-sections}

The differential cross section depends on amplitude as

\begin{equation}
 \frac{d\,\sigma}{d\,\Omega}=\frac{p\,m}{8\,\pi^2}\mid A\mid
 ^2.
\end{equation}

Then we have

\[
\left(\frac{d\,\sigma}{d\,\Omega}\right)=\frac{1}{3}\sum_{\sigma}\frac{2e^2\kappa
p\,\mid \vp\ve \mid^2} {(1-\kappa\,r_{t})(\kappa^2+p^2)^3}\,- \]

\[\frac{i}{3}\sum_{\sigma}\,\frac{e g
\overline{g}\,\kappa\,f(p)\,(\vp\ve^{*})\,[\vn\times\ve]}
{2\pi\sqrt{1-\kappa\,r_{t}}\,m\,(\kappa^2+p^2)^2}\langle\chi^{\sigma}\mid
2(\mu_{p}+\mu_{n})\,(\vI\,(\vp\vI)-
(\vp\vI)\vI)-i[\vp\times\vI]\mid\chi^{\sigma}\rangle.
\]

It is obviously, the parity violation contribution to the cross-section is
proportional to the average spin of the initial (unpolarized) deuteron. Thus the
$\pi$-meson exchange does not operate here. This fact for the total cross-section was
mentioned in the present paper \cite{korkh}.

 Now we are interesting in polarized
deuterons and unpolarized photons. Then after easy treatment one may obtain the total
expression for the differential cross-section

\begin{equation}
\left(\frac{d\,\sigma}{d\,\Omega}\right)_{tot}=\left(\frac{d\,\sigma}{d\,\Omega}\right)_{E1}+\Delta\frac{d\,\sigma}{d\,\Omega},
\end{equation}

where

\begin{equation}
\frac{d\,\sigma}{d\,\Omega}_{E1}=e^2\frac{\kappa\,p\,(p^2-(\vp\,\vn)^2)}{(1-\kappa\,r_{t})(\kappa^2+p^2)^3}
\end{equation}
is a ordinary expression for the electric dipole desintegration of the deuteron. And

\begin{equation}
 \Delta\frac{d\,\sigma}{d\,\Omega}=-\frac{\,e^2\,g\,\overline{g}\,\kappa\,p^2\,f(p)}{2\,\pi\,m\,\sqrt{1-\kappa\,r_{t}}\,(\kappa^2+p^2)^2}
 \left(\mu_{p}+\mu_{n}-\frac{1}{2}\right)\,\left(\vn\vI\, -\, \frac{\vI\vp\,\vn\vp}{p^2} \right)
\end{equation}
is a correction to the regular differential cross section due to the $\pi$-meson
exchange.

Now it is easy to integrate this result over angles and find the total cross-section

\begin{equation}
 \sigma=\frac{8\,\pi\,e^2\,\kappa\,p^3}{3\,(1-\kappa\,r_{t})(\kappa^2+p^2)^3}-\frac{4\,e^2\,g\,\overline{g}\,\kappa\,p^2\,f(p)}
 {3\,m\,\sqrt{1-\kappa\,r_{t}}\,(\kappa^2+p^2)^2}
 \left(\mu_{p}+\mu_{n}-\frac{1}{2}\right)\vn\vI.
\end{equation}

\section{Vector meson exchange}

Unfortunately, the $\pi$-meson contribution into the amplitude is not dominating.
There are also vector-meson (so called short-distance) contributions of two types. The
first of them is dominating near the threshold and the second one has the similar form
as the $\pi$-meson exchange.

These contributions may not be reliably calculated with good accuracy as it was
mentioned above. But we use the potential model to estimate their magnitudes. It is
important to understand the relation between cross-section correction due to the
$\pi$-meson exchange and the vector meson one.

We mean the Jastrov repulsion between nucleons at small distance
\cite{korkh},\cite{fts} to obtain wave function mixtures. According to the paper
\cite{mmr} we use perturbative deuteron wave function in the following form

\begin{equation}\label{adm1}
  \delta\psi_{d}=-i\lambda_{t}(\Si\,\gr)\sqrt{\frac{\kappa}{2\,\pi}}\frac{e^{-\kappa\,r}}{r}.
\end{equation}

Admixtures to the $s$ and $p$-waves of the continuous spectrum are

\[
\delta\psi_{s}=i\,\frac{\alpha_{s}}{1+i\,p\,\alpha_{s}}\,\lambda_{s}\Si\gr\left(\frac{e^{i\,p\,r}}{r}\right),
\]
\begin{eqnarray}\label{adm2}
\delta\psi_{p}=-\lambda_{s}\frac{\alpha_{s}}{1+i\,p\,\alpha_{s}}\Si\vp\,\frac{e^{i\,p\,r}}{r}.
\end{eqnarray}

Here $\alpha_{s}$ is a triplet scattering length, $\Si=\si_{p}-\si_{n}$, $\lambda_{t}$ and $\lambda_{s}$ are
\cite{korkh}

\[
\lambda_s= (0.028 h^0_{\rho} - 0.023 h^2_{\rho} + 0.028 h^0_{\omega})\times 10^{-7}
m_{\pi}^{-1}=-0.16\times 10^{-7} m_{\pi}^{-1}, \]

\begin{equation}\label{Lambd}
\lambda_t= (0.032 h^0_{\rho} + 0.001 h^0_{\omega})\times 10^{-7} m_{\pi}^{-1} = -0.37
\times 10^{-7} m_{\pi}^{-1}.
\end{equation}

Let us now consider the regular $E1$-transition from the deuteron into the $p$-wave of
the continuous spectrum. Due to wave function admixtures, the nonzero $M1$-transition
appears. It's straightforward calculation using the wave functions
(\ref{adm1},\ref{adm2}) give us the total amplitude

\[
A=-2\,e\,\sqrt{2\,\pi\,\omega}\sqrt{2\,\pi\,\kappa}\,\frac{\vp\,\ve}{\sqrt{1-\kappa\,r_{t}}(\kappa^2+p^2)^2}\delta_{\sigma'\sigma}-
i\frac{e}{2\,m}\,\sqrt{2\,\pi\,\omega}\sqrt{2\,\pi\,\omega}(\mu_{p}-\mu_{n})
\]
\begin{eqnarray}
\left(\frac{\lambda_{t}}{\kappa^2+p^2}\langle\chi^{\sigma'}\mid
\Si\left[\vn\times\ve\right]\,\Si\vp\mid\chi^{\sigma'}\rangle-
\frac{\alpha_{s}\,\lambda_{s}}{(1-i\,p\,\alpha_{s})(\kappa+i\,p)}
\langle\chi^{\sigma'}\mid\Si\vp\,\Si\left[\vn\times\ve\right]\mid\chi^{\sigma}\rangle\right).
\end{eqnarray}

If we are not interesting in photon's polarization effects we may average over all of
them. Then for the correction to the differential cross-section we have

\[
\Delta\frac{d\,\sigma}{d\,\Omega}=\frac{\kappa\,e^{2}\,(\mu_{p}-\mu_{n})p^3}{m\sqrt{1-\kappa\,r_{t}}\,(\kappa^2+p^2)^2}\,
\left[\left(\lambda_{t}+\alpha_{s}\frac{(\kappa+p^2\,\alpha_{s})}{1+p^2\,\alpha_{s}^2}\,\lambda_{s}\right)\,(\vI\vn-\frac{\vI\vp\,\vn\vp}{p^2})+\right.
\]
\begin{equation}
\left.\lambda_{s}\frac{\alpha_{s}}{p}\frac{1-\kappa\,\alpha_{s}}{1+p^2\,\alpha_{s}^2}
\left(\vp\vI\,[\vn\times \vp]\vI+\vI[\vn\times \vp]\,\vp\vI\right)\right]
\end{equation}
and for the total cross-section

\begin{equation}
\Delta\sigma=\frac{8\,\pi\kappa\,e^{2}\,(\mu_{p}-\mu_{n})p^3}{3
\,m\,\sqrt{1-\kappa\,r_{t}}\,(\kappa^2+p^2)^2}\,
\left(\lambda_{t}+\alpha_{s}\frac{(\kappa+p^2\,\alpha_{s})}{1+p^2\,\alpha_{s}^2}\,\lambda_{s}\right)\,\vI\vn.
\end{equation}

\section{Short-distance contribution near the threshold}

Now we consider the photodesintegration near the threshold. Here the magnetic dipole
transition $^{3}S_{1} \to ^{1}S_{0}$ is dominating. This transition does not conserved
the total spin. However, the total spin is conserved in the admixed $E1$-transition.
Therefore only the vector meson exchange (which does not conserve spin too) operates.

 All the calculations are very easy and the result for the amplitude is

 \[
 A=-i\frac{e}{2\,m}\sqrt{2\,\pi\,\omega}\sqrt{2\,\pi\,\kappa}(\mu_{p}-\mu_{n})
 \frac{1-\kappa\,\alpha_{s}}{(1-i\,p\,\alpha_{s})\,(\kappa^2+p^2)}\,\Si\left[\vn\times\ve\right]+
 \]
 \begin{eqnarray}
 \frac{e\,\sqrt{2\,\pi\,\omega}\sqrt{2\,\pi\,\kappa}}{3}(\lambda_{t}\,I_{t}+\lambda_{s}\,I_{s})\,\Si\ve.
 \end{eqnarray}

Here $I_{t}$ and $I_{s}$ are radial integrals

\[
I_{t}=\frac{3\kappa^2+p^2}{(\kappa^2+p^2)^2}-\frac{\alpha_{s}\,(2\kappa+i\,p)}{(1-i\,p\,\alpha_{s})\,(\kappa+i\,p)^2},
\]
\begin{eqnarray}
I_{s}=\frac{\kappa+2\,i\,p}{(1-i\,p\,\alpha_{s})\,(\kappa+i\,p)^2}.
\end{eqnarray}

One may obtain the differential cross-section after a simple algebra

\[
\frac{d\,\sigma}{d\,\Omega}=\frac{\kappa\,e^{2}\,(\mu_{p}-\mu_{n})^2\,(1-\kappa\,\alpha_{s})^2\,p}{4\,m^2\,(1+p^2\,\alpha_{s}^2)\,(\kappa^2+p^2)^2}\,(\vI\vn)^2+
\]
\[
\frac{\kappa\,e^{2}\,(\mu_{p}-\mu_{n})\,(1-\kappa\,\alpha_{s})\,p}{3\,m\,(1+p^2\,\alpha_{s}^2)}\,\frac{3\,\kappa^2+p^2}{\,(\kappa^2+p^2)^2}\,\times
\]
\begin{eqnarray}
(\lambda_{t}\,(1-\frac{2\,\alpha_{s}\,\kappa^3}{3\,\kappa^2+p^2})+\kappa\,\alpha_{s}\,\lambda_{s}\,\frac{\kappa^2+3\,p^2}{3\kappa^2+p^2})\vn\vI.
\end{eqnarray}

The total cross-section is

\[
\sigma=\frac{\pi\kappa\,e^{2}\,(\mu_{p}-\mu_{n})^2\,(1-\kappa\,\alpha_{s})^2\,p}{m^2\,(1+p^2\,\alpha_{s}^2)\,(\kappa^2+p^2)^2}\,(\vI\vn)^2+
\]
\[
\frac{4\pi\kappa\,e^{2}\,(\mu_{p}-\mu_{n})\,(1-\kappa\,\alpha_{s})\,p}{3\,m\,(1+p^2\,\alpha_{s}^2)}\,\frac{3\,\kappa^2+p^2}{\,(\kappa^2+p^2)^2}\,\times
\]
\begin{eqnarray}
(\lambda_{t}\,(1-\frac{2\,\alpha_{s}\,\kappa^3}{3\,\kappa^2+p^2})+\kappa\,\alpha_{s}\,\lambda_{s}\,\frac{\kappa^2+3\,p^2}{3\kappa^2+p^2})\vn\vI.
\end{eqnarray}

\section{Conclusion}

  We write the total cross-section as
\[
\sigma=\sigma_{E1}+\sigma_{M1}+\Delta\sigma_{\pi}+\Delta\sigma_{V1}+\Delta\sigma_{V2},
\]
\[
 \sigma_{E1}=\frac{8\,\pi\,e^2\,\kappa\,p^3}{3\,(1-\kappa\,r_{t})(\kappa^2+p^2)^3},
\]
\[
\sigma_{M1}=\frac{\pi\kappa\,e^{2}\,(\mu_{p}-\mu_{n})^2\,(1-\kappa\,\alpha_{s})^2\,p}
{m^2\,(1+p^2\,\alpha_{s}^2)\,(\kappa^2+p^2)}\,(\vI\vn)^2,
\]
\[
\Delta\sigma_{\pi}=\frac{4\,e^2\,g\,\overline{g}\,\kappa\,p^2\,f(p)}
 {3\,m\,\sqrt{1-\kappa\,r_{t}}\,(\kappa^2+p^2)^2}
 (\mu_{p}+\mu_{n}-\frac{1}{2})\vI\vn,
\]
\[
\Delta\sigma_{V1}=\frac{8\,\pi\kappa\,e^{2}\,(\mu_{p}-\mu_{n})p^3}{3
\,m\,\sqrt{1-\kappa\,r_{t}}\,(\kappa^2+p^2)^2}\,
(\lambda_{t}+\alpha_{s}\frac{(\kappa+p^2\,\alpha_{s})}{1+p^2\,\alpha_{s}^2}\,\lambda_{s})\,\vI\vn,
\]
\[
\Delta\sigma_{V2}=\frac{4\pi\kappa\,e^{2}\,(\mu_{p}-\mu_{n})\,(1-\kappa\,\alpha_{s})\,p}{3\,m\,(1+p^2\,\alpha_{s}^2)}\,\frac{3\,\kappa^2+p^2}{\,(\kappa^2+p^2)^2}\,
\left(\lambda_{t}\,(1-\frac{2\,\alpha_{s}\,\kappa^3}{3\,\kappa^2+p^2})+\kappa\,\alpha_{s}\,\lambda_{s}\,\frac{\kappa^2+3\,p^2}{3\kappa^2+p^2}\right)\vI\vn.
\]

\begin{figure}
 \centering\includegraphics[height=10cm]{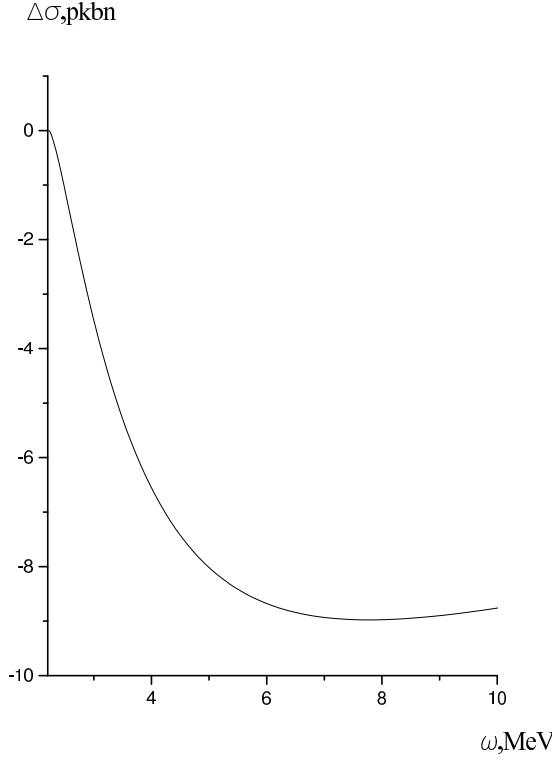}
 \caption{$E1$ regular transition. The $\pi$-meson contribution into $\Delta\sigma$.}
\end{figure}

\begin{figure}

\centering\includegraphics[height=10cm]{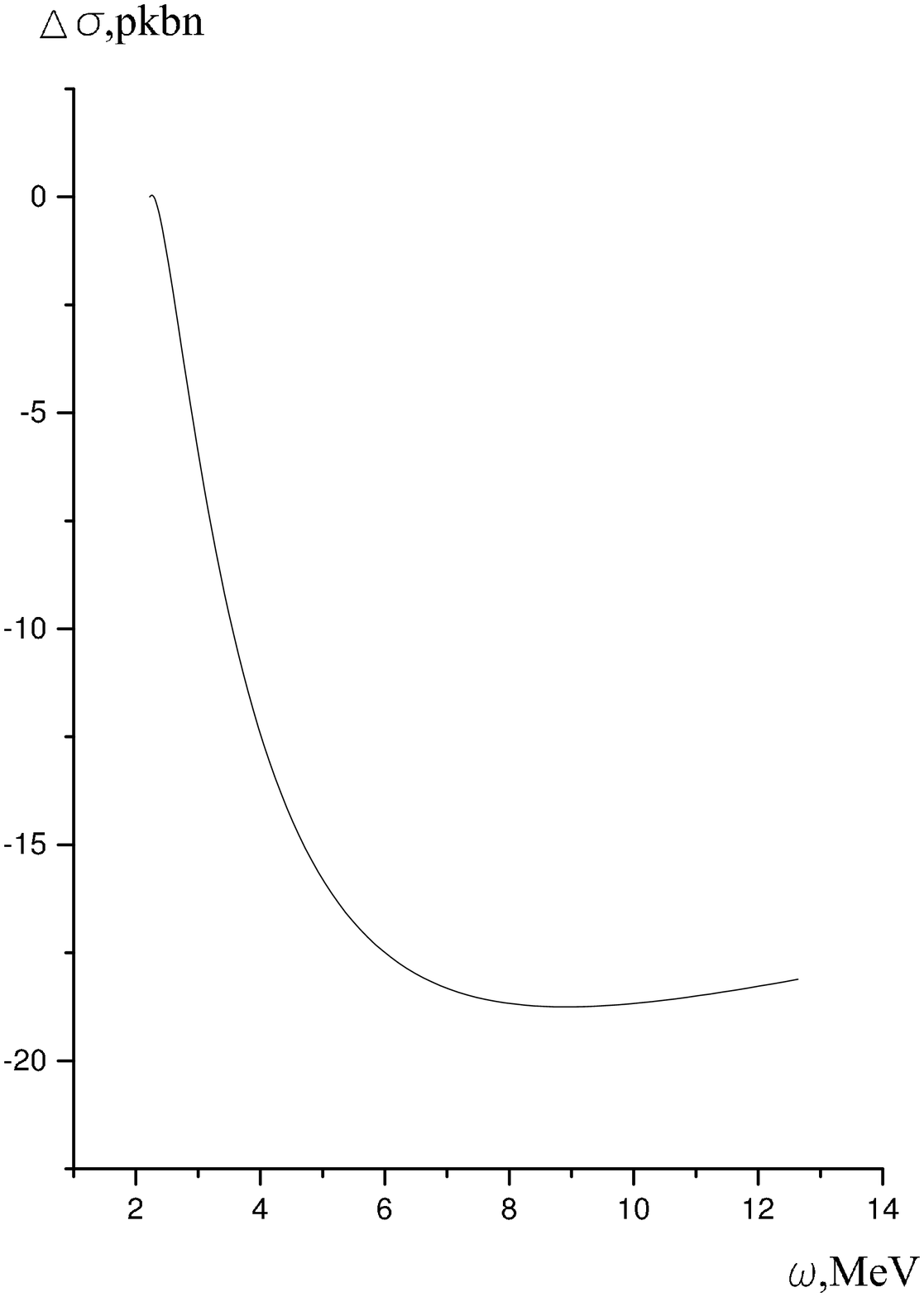}        
\caption{$E1$ regular transition. The vector-meson contribution into $\Delta\sigma$.}
\end{figure}

\begin{figure}

\centering\includegraphics[height=10cm]{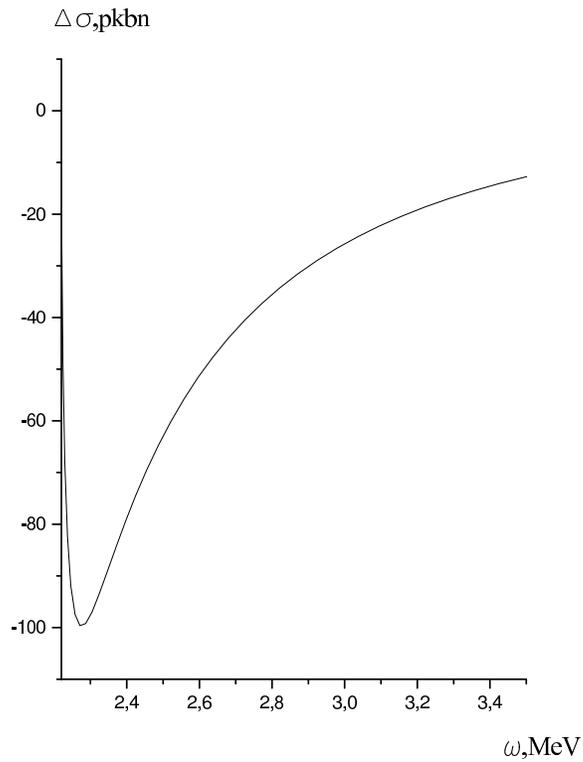}        
\caption{$M1$ regular transition. The vector-meson contribution into $\Delta\sigma.$}
\end{figure}

  Let us estimate now results and their precisions. We accord to the "so called" best
values of weak constants (supported by the experimental result for the $^{133}$Cs
anapole moment \cite{wi}). Then the weak $\pi N N$ constant is

\[
\overline{g}=3.3\, \times\, 10^{-7}.
\]

Correction to the total cross-sections (there $\vn\vI=1$) are plotted in figures. Let
us talk now about the $\pi$-meson (Fig.1) and the vector-meson (Fig.2) contributions
to the regular $E1$-transition. Unfortunately, the vector meson contribution is
dominating here and it is more than the $\pi$-meson one of a factor 2 in the main
region of energies. The accuracy of the vector meson exchange is too bad as it was
mentioned above  because of the unreliable calculations at the small range. But the
main error we have in a calculation of $\lambda_{t},\,\lambda_{s}$ constants. Our
calculations of these constants with and without Jastrow repulsion discrepancies of
the factor less than 1.7. So, we may expect that the accuracy of this result is about
$40\%$ for given parameters of weak constants. The precision  of the $\pi$-meson one
is $20\%$. The last estimate one may obtain via comparison of two results of the
$\pi$-meson contribution with the zero range approximation function and with model
function \cite{kk}.
 We obtain that the magnitude of
$\Delta\sigma_{\pi}/\sigma$ is about $0.3\times 10^{-9}$ that is essentially less than
the maximal magnitude due to the vector meson exchange.

 The
maximal magnitude has the vector meson contribution to the magnetic dipole regular
transition (Fig.3), which relative magnitude  $\Delta\sigma/\sigma$ is about $4\times
10^{-8}$.  The accuracy is again 40\%.

\section{Acknowledgements}
I very much appreciate to Khriplovich I.B. for helpful discussions and useful notices
and to Krupnikov E.D. for careful reading the paper.
 The work was supported by
the Russian Foundation for basic Research through Grant No. 01-02-16898, No.
00-15-96811 and by Ministry of Education through Grant No. …00-3.3-148.

\end{document}